# Fermi arcs as a visible manifestation of pair level of negative-U centers

Kirill Mitsen, Olga Ivanenko

*Lebedev Physical Institute, Leninskii pr., 53, 119991, Moscow, Russia*

**Abstract**

Here we consider the possibility that the Fermi arcs observed in ARPES experiments result from interaction of valence band electrons with negative-U centers (NUCs) which are formed under doping on the pairs of neighboring Cu cations in $CuO_2$ plane. This interaction results in two-particle hybridization and vanishing of the gap over the part of Fermi contour. Furthermore the transitions of electrons to NUCs result in the generation of free hole carriers. Just these carriers rather than doping induced charges provide conduction in the normal state.

___

Earlier [1] we proposed the mechanism of negative-U centers (NUCs) formation in high-$T_c$ superconductors. The model is based on the assumption of rigid localization of doped charges in the close vicinity of doped ion. The localization results in local variation of electronic structure of the parent charge-transfer insulator. This variation depends on the local mutual arrangement of the doped charges. There was shown that NUCs may be formed under certain conditions on the pairs of neighboring Cu cations in $CuO_2$ plane.

Figure 1a shows the electronic spectrum of undoped high-$T_c$ superconductor. Here the charge-transfer gap $\Delta_{ct}$ corresponds to the transition of electron from oxygen to nearest Cu ion. The appearing hole is extended over 4 surrounding oxygen ions (Fig. 1b) due to overlapping of their $2p_{x,y}$ orbitals. This exitation ($3d^{10}$-electron+2p-hole) resembles a hydrogen atom. We have shown that the energy of two such excitations can be lowered (Fig. 1c) if two side-by-side pseudo-atoms form a pseudo-molecule (Fig. 1d). This is possible due to formation of a bound state (of the Heitler-London type) of two electrons and two holes that emerge in the immediate vicinity of this pair of Cu ions. It implies the formation of NUC on the pair of neighboring Cu ions.

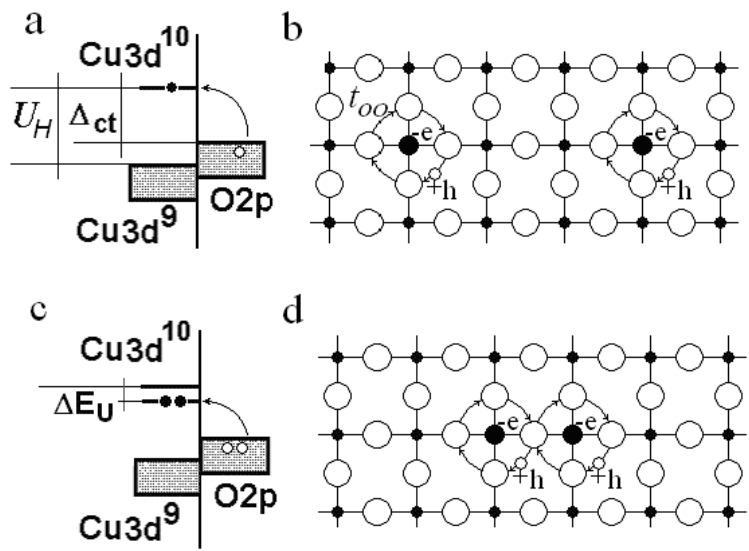

Fig. 1. (a) Electronic spectrum of undoped cuprate high-$T_c$ superconductor. $U_H$ is the repulsive energy for two electrons on Cu ion. The charge transfer gap $\Delta_{ct}$ corresponds to the transition of electron from oxygen to nearest Cu ion with the origin of hole extended over four surrounding oxygen ions (b). (c) The energy of two such excitations can be lowered if two side-by-side pseudo-atoms form pseudo-molecule (d).

If we decrease $\Delta_{ct}$ somehow to the point where the gap disappears for two-particle transitions but survives for one-particle transitions ($\Delta_{ct}=\Delta'$), we arrive at a system in which electrons belonging to the valence band effectively interact with pair states of NUCs. Such NUCs we term as active NUC. We believe that the role of the doping is just to suppress the gap $\Delta_{ct}$ down to $\Delta'$ and to activate NUC's. At the determinate doping each pair of Cu ions is active NUC.

Hybridization of pair states of NUC and one-electron band states results in building of half-filled (taking account of NUC pair states) band but with pairing of electrons ($k,-k$) near $k_F$ due to renormalization electron-electron interaction because of virtual transitions of electron pairs to NUC pair level. Hybridization and electron pairing result in an electronic spectrum in which the pair electronic level of NUC $E_p$ is placed at the top of subband of coupled states in the vicinity of points ($\pm\pi/2;\pm\pi/2$) (Fig. 2a). The fulfilled subband is separated from empty upper Habbard band (band of one particle excitations) with the gap $\Delta'$. The pairing potential $\Delta$ is momentum-dependent, vanishing in a node direction (i.e., for the electrons that move only in the oxygen sublattice) and reaching a maximum $\Delta$ in the antinode (copper–oxygen) direction (i.e., for the electrons that have the maximum frequency of transitions to the NUC).

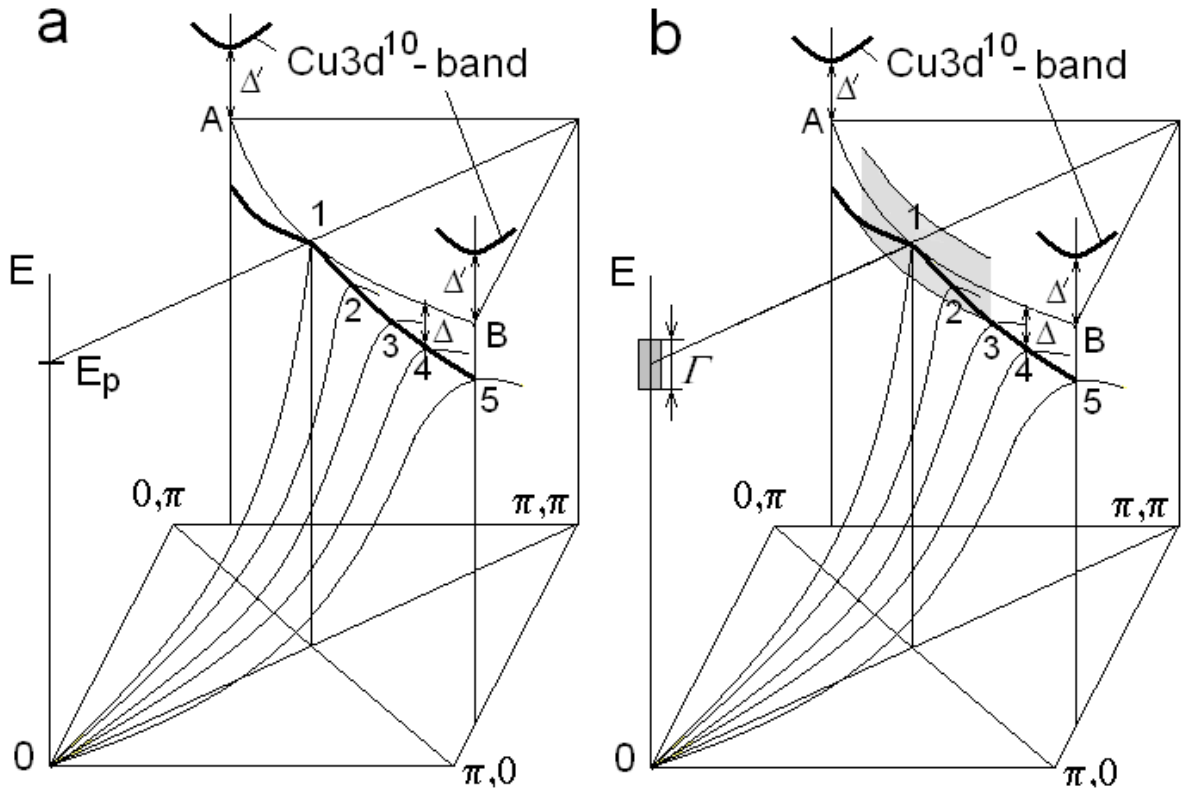

Fig. 2. The development of Fermi arc in cuprate HTSC as result of interaction of valence band electrons with pair states of NUCs. a) $T=0$, b) $T>0$. The curves *01-05* are the curves of quasiparticle dispersion. *A1B* – Fermi contour. $E_P$ and $\Gamma$ – energy and width of NUC pair level. Shaded strip around $E_p$ is the region of moments of electronic pairs (***k***, -***k***) occupying NUC's. Solid curve (1-5) is the upper boundary of occupied states.

Noncoherent transport of bound electron pairs in this system is impossible. However at $T>0$ pair excitations which correspond to the transition of two electrons to the pair level with formation of two holes in a fulfilled subband are possible. Owing to orbital overlapping these holes (with concentration $n(T)$) can move through a crystal and just these holes provide the normal conductivity. The occupation of NUC with real electrons hinders the development of superconducting coherency in the system of electrons pairs due to decreasing of number of states available for their scattering. The superconducting coherency will take place at the temperature at which the NUC occupation $\eta$ will less than some critical value $\eta_c$ [1].

Let consider the dependences of $n$ and $\eta$ on temperature. The lifetime of the pair state of the NUC is determined by the pair-level width $\Gamma$ (Fig. 2); subject to the two-particle hybridization, it is [2, 3]

$$\Gamma \approx kT(V/E_F)^2 \qquad (1)$$

where $V \sim 1$ eV is the one-particle p–d hybridization constant, $E_F \sim 0.4$ eV is Fermi energy [4], and $T$ is the temperature. Thus, the broadening of the pair level is $\Gamma \sim 5kT$. The corresponding broadening of the band states $\gamma$ is also proportional to $T$ [2].

The interaction of band electrons with NUC results in infill of the gap with states provided $2\Delta<\Gamma$. As the temperature increases the gapless region at the Fermi level extends from the point ($\pi/2$, $\pi/2$) toward a low dispersion, forming an arc of thickness $\Gamma \sim 5kT$ and of length $l \propto T$ along the boundary of the remnant Fermi surface [4].

As noted above active NUCs play the role of pair acceptors and lead to the generation of free hole carriers in the $CuO_2$ plane. If $N_U$ is the concentration of active NUCs and $\eta$ is their average population ($0<\eta<2$), the concentration of the forming hole carriers is $n=\eta N_U$. As follows from the model, these carriers rather than doped charges (that localized in the close vicinity of doped ion) provide conductivity in the normal state. In turn $n(T)$ is determined by the equality of the rates of forward and reverse pair level–band transitions. The pair level–band transition frequency is $\eta\Gamma \propto T\eta$. The rate of the reverse process is specified by the number of electron pair states ($k,-k$) available for transitions to NUC from energy interval $\sim\Gamma$ that is proportional to $T^2$ (Fig. 2a) and by the occupation of NUC $\eta$. Therefore the rate of the band–pair level transitions is proportional to $T^2(2-\eta)$. Thus, $\eta T \propto (2-\eta)T^2$, and we have

$$\eta=2T/(T+T_0) \text{ and } n=2N_UT/(T+T_0) \qquad (2)$$

$T_0 \approx 390K$ is temperature independent constant determined from comparison with experiment [5]. The critical value $\eta_c$ can be estimated as $\eta_c=2T_c/(T_c+T_0) \approx 2/5$.

The pair level–band transition frequency, which is $\eta\Gamma$, is also the frequency of hole–hole scattering $v_{pp}$ with a transition of hole pairs to a NUC. Given $T_0$ and the hole–hole scattering frequency $v_{pp}=2\Gamma T/(T+T_0)$ we can estimate the electrical resistivity for the case of optimum doping, when every cell has a NUC, using the Drude formula $\rho=mv_{pp}/ne^2$. For this case, our estimate is $\rho(100K)\sim 50\mu\Omega\cdot$cm which agrees with the results of numerous experiments.

This research was supported by the Russian Foundation for Basic Research (grant 08-02-00881)